\documentclass[draftcls,onecolumn,12pt,journal]{IEEEtran}
\usepackage{amsmath, amssymb,amsthm}
\usepackage{graphicx, float, adjustbox, multirow, authblk, enumerate, array, hyperref, siunitx}
\usepackage[caption=false, font=footnotesize, labelfont=sf,textfont=sf]{subfig}
\usepackage[numbers,sort,square,compress]{natbib}

\usepackage{balance}

\usepackage{cuted}
\usepackage{algorithm,algpseudocode}

\newcommand{\MC}[1]{\mathcal{#1}}

\newcommand{\e}[1]{\mathrm{e}^{#1}}

\newcommand{\imgwidth}{0.65\columnwidth}

\newcommand{\sref}[1]{\protect\subref{#1}}
\renewcommand{\vec}[1]{{\mathbf{#1}}}
\newcommand{\vecsym}[1]{{\boldsymbol{#1}}}
\newcommand{\K}{\vecsym{\mathcal{K}}}
\newcommand{\T}{\vecsym{\mathcal{T}}}
\newtheorem{theorem}{Theorem}
\title{Lattice All-Pass Filter based Precoder Adaptation for MIMO
  Wireless Channels}
\author{Parth Mehta$^{\dagger}$, Agulla Surya Bharath, Kumar Appaiah~\IEEEmembership{Member,~IEEE}, Rajbabu Velmurugan~\IEEEmembership{Member,~IEEE}, Debasattam Pal~\IEEEmembership{Member,~IEEE}
\thanks{~}\thanks{Manuscript created \today.}}

\begin{document}
\maketitle
\begin{abstract}
Modern 5G communication systems employ multiple-input multiple-output
(MIMO) in conjunction with orthogonal frequency division multiplexing
(OFDM) to enhance data rates, particularly for wideband millimetre
wave (mmW) applications. Since these systems use a large number of
subcarriers, feeding back the estimated precoder for even a subset of
subcarriers from the receiver to the transmitter is
prohibitive. Moreover, such frequency domain approaches do not
exploit the predominant line-of-sight component present in
such channels to reduce feedback. In this work, we view the precoder
in the time-domain as a matrix all-pass filter, and model the
discrete-time precoder filter using a matrix-lattice structure that
aids in reducing the overall feedback while maintaining the
desired frequency-phase delay profile. This provides an efficient
precoder representation across the subcarriers using fewer
coefficients, and is amenable to tracking over time with much lower
feedback than past approaches. Compared to frequency domain Geodesic
interpolation, Givens rotation parameterisation, and the
angle-delay domain approach that depends on approximate discrete-time
representation, the proposed approach yields higher achievable rates
with a much lower feedback burden. Via extensive simulations over mmW
channel models, we confirm the effectiveness of our claims, and show
that the proposed approach reduces the feedback burden by up to
70\%.
\end{abstract}

\section{Introduction}
\label{sec:intro}
Multiple-input multiple-output (MIMO) based wireless communication
systems have become the mainstay of several modern wireless
communication systems. In conjunction with orthogonal frequency
division multiplexing (ODFM), MIMO has enabled the development of
scalable high-speed wireless systems. The 5G and beyond 5G standards
that are targeted to meet enhanced data rate requirements, largely use
high-frequency carrier (millimetre wave (mmW), terahertz (THz) etc.)
with wideband OFDM and larger antenna arrays (massive MIMO), along
with smart algorithms to increase
degrees-of-freedom~\cite{albreem2021overview,domouchtsidis2020constant,
yao2018semidefinite}. An
essential ingredient in achieving high rates over such links is to use
precoding to direct signals effectively in the spatial
domain. Effective precoding across all subcarriers for such a system
requires channel state information (CSI) to be made available at the
transmitter, typically through a limited feedback link from the
receiver. Nearly, all current feedback techniques for these systems
use frequency domain CSI, which places a large burden on the
feedback channel in the wideband MIMO scenario that is fast becoming
the norm in modern wireless communication systems. In this paper, we
discuss a more structured feedback approach that uses efficient
time-domain realisation of matrix all-pass filters to significantly
reduce the feedback requirement without compromising achievable
rates. We further show that exploiting temporal correlations of the
time varying channel enables tracking of the precoders across all
subcarriers with very little incremental feedback.

Precoding in MIMO systems has been studied significantly in the past,
and is also part of several standards. Because of the larger array
size, we restrict ourselves to the linear precoding schemes, since
they are advantageous over non-linear methods in terms of performance
and computation cost. In \cite{fatema2017massive}, a comprehensive
survey of linear precoding techniques for massive MIMO
systems is covered. Several precoding techniques such as Maximum Ratio
Transmission (MRT), Zero-Forcing (ZF), Minimum Mean Square Error
(MMSE), Truncated Polynomial Expansion (TPE), matrix decomposition
based methods (like QR decomposition and singular value decomposition
(SVD)) influence precoder choice and
design~\cite{albreem2021overview}. State-of-the-art approaches to
precoder design, targeted for 5G related MIMO applications. Precoder
design and use in established 5G channel models like 3GPP SCM,
WINNER-II that are suitable for specific scenarios are discussed
in~\cite{baum2005interim,bultitude20074,mondal20153d,domouchtsidis2020constant,
ghirmai2016design,guo2016robust,yao2018semidefinite,gupta2018optimised,
taghizadeh2017linear}.
More generalised models that incorporate massive MIMO, V2V, HST and
mmW small-scale fading models for 5G and beyond 5G applications have
also seen research interest
lately~\cite{wu2017general,bian2021general}.

Typically, the frequency division duplex (FDD) based systems consider
the channel matrix in the frequency domain for processing, and the
precoding is done for every subcarrier separately. A major drawback
here would be huge burden on the receiver to feed back the full or
partial CSI across a multitude of subcarriers in wideband
communication systems. Since feeding back precoders for all
subcarriers is restrictive, particularly for wideband systems, most
practical deployments feed back the precoder only for certain
subcarriers called pilot subcarriers, while the remaining precoders are
interpolated. Interpolating precoders typically exploit the unitary
structure of the precoder, and use either manifold based
techniques~\cite{nijhawan2021flag,love2003grassmannian}, or
parameterisation of the precoders and interpolation of these
parameters~\cite{godana2013parametrization}. All these approaches do
not necessarily exploit any potential structure that the realisation
of the precoder as a filter [in the form of a linear constant
coefficient difference equation (LCCDE)] may offer. The concept of
time-domain precoding for SISO systems has been considered in the
past~\cite{liu2018time}, and this has further led to discussions on
time-domain precoding for MIMO
systems~\cite{solomennikova2021frequency}, wherein it is suggested
that the channel impulse response $\vec{H}[n]$ be computed
from the frequency-domain channel
matrix $\vec{H}(\e{j\omega})$, and and only a few non-zero
taps of this impulse response that have comparatively higher channel
gain be used to compute the precoding matrix. However, it is yet to be
established that time-domain approach offer improved performance when
compared to the conventional frequency domain based methods. There are
several 5G and beyond 5G channels, wherein the precoder, as a function
of frequency, exhibits a structure that can be better captured if
expressed in the time-domain, thus leading to improved
quantisation. We expand on this approach, with the following
contributions:
\begin{itemize}
    \item We extensively use the lattice based matrix all-pass filter
    (time-domain) representation to capture the precoders for all
    subcarriers, and show that the time domain realisation is able to
    capture the precoders' characteristics using very few
    coefficients. The realisation of lattice based matrix all-pass
    filters is well known for the case of real
    coefficients~\cite{regalia1988digital, vaidyanathan1985general,
    vaidyanathan1989role,vaidyanathan1987unified}, and we extend this
    to the case of complex coefficient based filters, since this is
    necessary to capture precoders for complex baseband channels.We
    compute the linear constant-coefficient difference equation
    (LCCDE) form using subspace Nevanlinna-Pick interpolation (SNIP)
    approach~\cite{unpublishedkeyS} and then convert them to lattice
    coefficients.

    \item We show that in the lattice approach to representing the
    time-domain precoder filters, adapting the precoders involves
    tracking the lattice parameters that are amenable to easy
    tracking, while also ensuring the stability of the realised
    filter. This also has the distinct advantage of capturing the
    precoder characteristics across the complete bandwidth of
    operation, unlike other approaches that capture the approximate
    filter characteristics using only only a few dominant-path
    components~\cite{solomennikova2021frequency,mo2017channel}.

    \item Through extensive simulations, we verify that the proposed approach is effective
    in capturing the wide band precoder, and is amenable to effective tracking over
    time to yield high data rates. For several 5G channel models, the proposed method
    is comparable to the traditional frequency-domain based quantisation.
\end{itemize}

The rest of this paper is organised as follows: Section~\ref{sec:sys} shows the formulation
of channel model, channel adaptation, and method to calculate achievable rates for a given
precoding scheme. Section~\ref{sec:matrixfilter} explains how to realise the precoder matrix
for the wireless MIMO channel model in LCCDE form, and convert it into lattice structure, as
well as the advantages of the matrix lattice realisation in terms of stability, quantisation,
and adaptability. Section~\ref{sec:results} shows the quantitative comparison of achievable
rates between the geodesic~\cite{khaled2005interpolation}, Givens rotation~\cite{roh2007efficient,
madan2020scalar} and lattice implementations for various time-varying channel models. Finally,
Section~\ref{sec:conclude} summarises and concludes the proposed method.

\section{System Model and Achievable Rates}
\label{sec:sys}
\subsection{Channel model}
\label{subsec:model}
We consider a typical MIMO-OFDM discrete-time baseband channel model,
as below:
\begin{equation}
    \vec{y}_t(\e{j\omega})
    = \vec{H}_t(\e{j\omega}) \vec{x}_t(\e{j\omega})
    + \vec{w}_t(\e{j\omega}), \quad \omega \in (-\pi, \pi]
    \label{eq1:sysmodel}
\end{equation}
where $\vec{x}_t$ and $\vec{y}_t$ are the transmitted and received signals, respectively,
$\vec{H}_t$ is the complex channel matrix, and $\vec{w}_t$ is additive white Gaussian noise.
Note that all the components mentioned here are functions of frequency $\omega$, and varying
with time $t$. Denoting the $k$th sub-carrier frequency by $\omega_k$,
\begin{equation}
    \vec{y}_t[k] = \vec{H}_t[k] \vec{x}_t[k] + \vec{w}_t[k]
    \label{eq6:sysmodel}
\end{equation}

Equivalently in the time-domain,
\begin{equation}
    \tilde{\vec{y}}_t[n] = \tilde{\vec{H}}_t[n] * \tilde{\vec{x}}_t[n] + \tilde{\vec{w}}_t[n]
    \label{eq2:sysmodel}
\end{equation}
where $*$ denotes convolution. The channel
matrix $\vec{H}_t$ is of size $N_R \times N_T$,
where $N_T$ and $N_R$ represents number of transmit
and receive
antennas. Here, $\vec{x}_t(\e{j\omega}),~\vec{y}_t(\e{j\omega}),~\vec{H}_t(\e{j\omega})$
are discrete-time Fourier transforms
of $\tilde{\vec{x}}_t[n],~\tilde{\vec{y}}_t[n],~\tilde{\vec{H}}_t[n]$
at any given time $t$, respectively. We compute the frequency
response of such a channel $\vec{H}_t(\e{j\omega})$ from a
power-delay profile. Given the matrix
taps $[\vec{H}_t[0],~\vec{H}_t[1],~\dots,~\vec{H}_t[L]]$ and
the relative
delay $[\tau_{t_0},~\tau_{t_1},~\dots,~\tau_{t_L}]$ of each
component, $\vec{H}_t(\e{j\omega})$ can be obtained as
\begin{equation}
    \vec{H}_t(\e{j\omega}) = \sum\limits_{l=0}^{L}\vec{H}_t[l] \e{-j\omega \tau_{t_l}}.
    \label{eq5:sysmodel}
\end{equation}

The channel $\vec{H}_t$ considered here is slowly varying, and is modelled using an AR(1)
process as shown in \eqref{eq3:sysmodel}~\cite{kim2011mimo}. Each element $\left[\vec{H}_{t}\right]_{ij}$
in the channel matrix $\vec{H}_t$ is modelled as
\begin{equation}
    \left[\vec{H}_{t}[n]\right]_{ij} = \alpha_{ij} \left[\vec{H}_{t-1}[n]\right]_{ij} + \sqrt{1-\alpha_{ij}^2} \left[\vec{W}_{t}[n]\right]_{ij}
    \label{eq3:sysmodel}
\end{equation}
where $\vec{H}_{t-1}$ is the channel at previous time instant, $\left[\vec{W}_{t}\right]_{ij} \sim \MC{N}(0,1)$, and $\alpha_{ij} \in (0,1)$. The value of $\alpha$ is chosen to reflect the rate of channel variation, as in \eqref{eq4:sysmodel}~\cite{kim2011mimo}.
\begin{equation}
    \alpha = J_0(2\pi f_d T_s)
    \label{eq4:sysmodel}
\end{equation}
where $J_0(.)$ is the 0th order Bessel function, $f_d = \frac{v}{c}F_c$ is the relative Doppler
frequency, given the carrier frequency $F_c$, speed $v$ and the speed of electromagnetic waves
in free-space $c$; and $T_s$ is the symbol duration.

\subsection{Optimal Precoding and Achievable Rates}
\label{secsec:prerate}
The transmit vector $\vec{x}_t$ in a MIMO-OFDM setup is typically precoded. The purpose of
this precoder is to maximise the overall achievable rate for a given channel $\vec{H}_t$.
Denoting the precoder as $\vec{P}_t$, the transmit vector for the $k$th subcarrier can be given as
\begin{equation}
    \vec{x}_t[k] = \vec{P}_t[k]\vec{d}_t[k], ~k = 0,~1,~\dots,~N_{\text{FFT}}-1
    \label{eq1:sysprerate}
\end{equation}
where $\vec{d}_t$ is the data, and $N_{\text{FFT}}$ is the number of subcarriers.

\begin{figure}[hbt]
    \centering
    \includegraphics[width=\imgwidth]{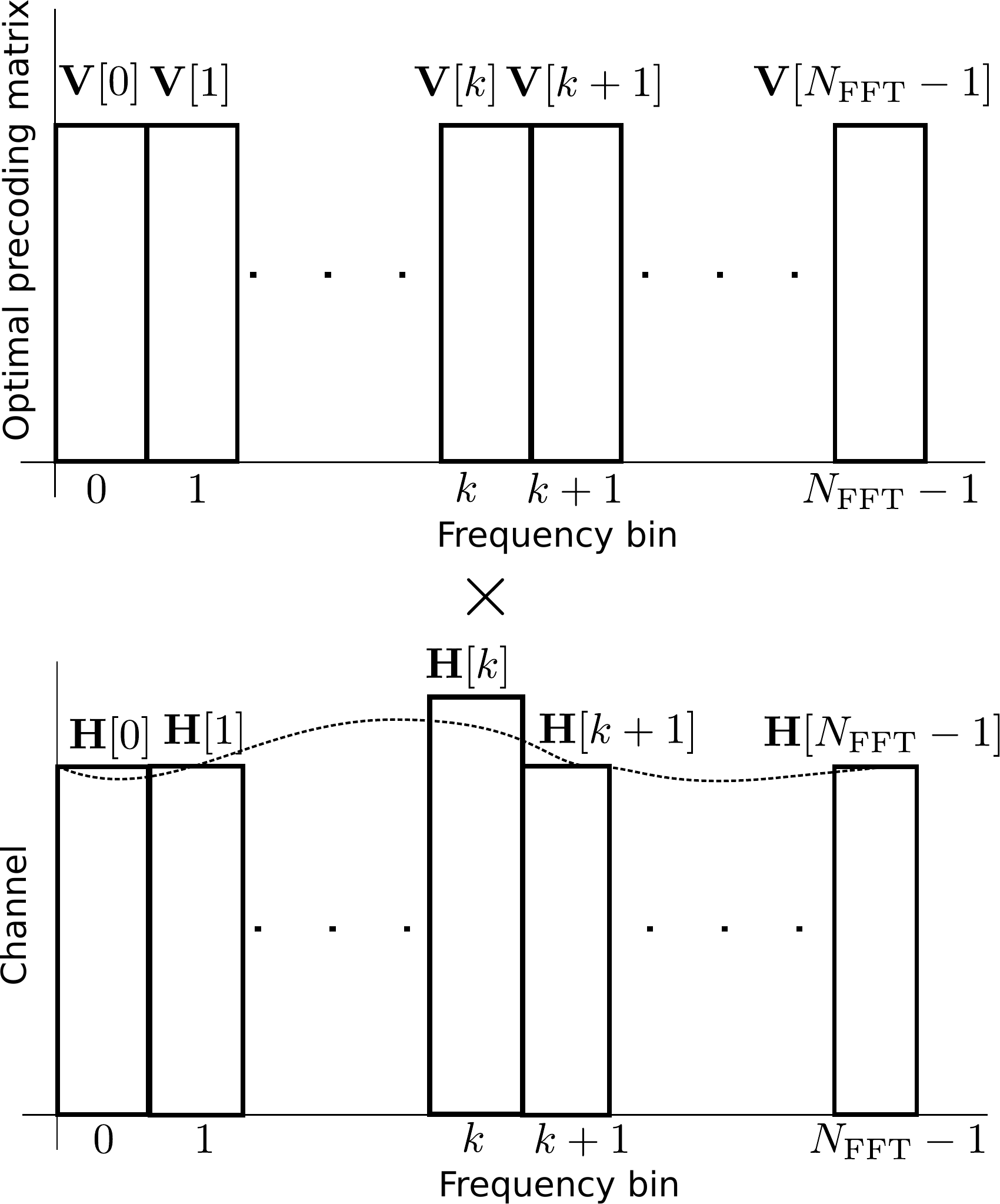}
    \caption{Abstract representation of approximated magnitudes of OFDM subcarriers
    for $N_{\text{FFT}}$ subcarriers.}
    \label{fig:ofdm_sub}
\end{figure}

To obtain the optimal achievable rate for a linear receiver, the
precoder $\vec{P}_t$ is the matrix of right singular vectors
of the channel matrix $\vec{H}_t$~\cite{pitaval2013codebooks}. Hence, for every
subcarrier $k$, $\vec{P}_t[k] = \vec{V}_t[k]$, and
$\vec{H}_t[k] =
  \vec{U}_t[k]\vecsym{\Sigma}_t[k]\vec{V}^H_t[k]$ as shown in
\figurename~\ref{fig:ofdm_sub}, where $\vecsym{\Sigma}_t[k]$
is a diagonal matrix consisting of the singular values of $\vec{H}_t[k]$. This precoder
is estimated at the receiver, and fed back to the transmitter for all subcarriers.

Now, given the channel $\vec{H}_t$ and precoder $\vec{P}_t$, the equivalent channel
is $\vec{H}_{t_{\text{eq}}} = \vec{H}_t\vec{P}_t$. With the zero-forcing (ZF) linear
receiver~\cite{love2005multimode,pitaval2013codebooks} of the form $\vec{Z} = \vec{F}\vec{H}_{t_{\text{eq}}}^H$,
where $\vec{F} = (\gamma\vec{H}_{t_{\text{eq}}}^H\vec{H}_{t_{\text{eq}}})^{-1}$,
the achievable rate can be computed as
\begin{equation}
    r_{\text{ZF}} = \sum\limits_{k=0}^{N-1} \log_2{\left(1 + \frac{1}{\left[\vec{F}\right]_{kk}}\right)}
    \label{eq3:sysprerate}
\end{equation}
where $N$ is the number of streams (number of diagonal elements of $\vec{F}$),
and $\left[\vec{F}\right]_{kk}$ are the diagonal elements of $\vec{F}$~\cite{pitaval2013codebooks}.

\subsection{Precoder Adaptation}
\label{subsec:preadapt}
To track the precoder for slowly varying channels shown in
Section~\ref{subsec:model}, adaptive quantisation approaches such as
Adaptive-Delta Pulse-Coded Modulation (ADPCM) are typically employed, since
they track only the precoder changes, and thus require very little
feedback. As shown in \cite{roh2007efficient,madan2020scalar}, to
track small changes in a quantity ($\hat{x}$), only 1-bit
is sufficient to determine the direction of the update (i.e., positive
or negative) of the parameter. This can be done by tracking the
sign-bit $\beta$. For frequency based methods, this has to be
done for every subcarrier (or every pilot subcarrier, if the remaining are
interpolated). The approach to track the time-varying parameter $x_t$
using single bit updates at each time instant can be represented as follows:
\begin{equation}
    \hat{x}_t = \hat{x}_{t-1} + \beta_t \mu_t,~\text{where}~\mu_t = \begin{cases} \sigma\mu_{t-1}&: \beta_t = \beta_{t-1} \\ \mu_{t-1}/\sigma&: \beta_t \neq \beta_{t-1}\end{cases}
    \label{eq1:syspreadapt}
\end{equation}
where $\sigma>1$ is the step size.

In the case of a MIMO-OFDM system that contains a large number of
antennas as well as a large number of subcarriers, the amount of data
that the receiver has to feedback to the transmitter is large, even
when adaptive feedback is employed. One approach to
reduce the amount of feedback is for the receiver to track and feed
back precoder updates only for pilot subcarriers, so that the transmitter
can interpolate to obtain the precoders at the remaining
subcarriers~\cite{khaled2005interpolation}. This method, while effective in several
situations, is highly sensitive to the number of pilot subcarriers. This can be
attributed to the fact that frequency domain interpolation does not guarantee
that the characteristics are satisfied for any frequencies
other than the pivot points that are used for interpolation. This
issue is exacerbated in the case of matrix filters, since
interpolation accuracy suffers significantly when using matrix based filters.

One approach to address this issue is to parameterise the unitary
matrices that represent the precoders as scalar
parameters using Givens rotations~\cite{roh2007efficient}. This
approach yields a unique  set of scalar parameters that completely
capture the characteristics of unitary matrices (such as the precoding
matrices) using a set of parameters $\phi_{k,l}$ and
$\theta_{k,l}$, as shown in \eqref{eq2:syspreadapt}. A
unitary matrix $\vec{V}_t$ of size $m \times m$ can
be decomposed a follows:
\begin{equation}
    \vec{V}_t(\e{j\omega}) = \left[ \prod_{k=1}^{m} \vec{S}_k(\phi_{k,k},\dots,\phi_{k,m}) \prod_{l=1}^{k} \vec{C}_{m-l,m-l+1}(\theta_{k,l}) \right],~\forall \omega.
    \label{eq2:syspreadapt}
\end{equation}
Here, $\vec{S}_k(\phi_{k,k},\dots,\phi_{k,m}) = \left[ \text{diag}(\vec{1}_{k-1},\e{j\phi_{k,k}},\dots,\e{j\phi_{k,m}}) \right]$
with $\vec{1}_k$ is $k-1$ ones; and
\begin{equation*}
    \vec{C}_{m-l,m-l+1}(\theta) = \begin{bmatrix} \vec{I}_{m-l-1} & ~ & ~ & ~ \\ ~ & \cos{\theta} & -\sin{\theta} & ~ \\ ~ & \sin{\theta} & \cos{\theta} & ~ \\ ~ & ~ & ~ & \vec{I}_{l-1}\end{bmatrix}
  \end{equation*}
where the blank entries are all zeros. $\phi_{k,l}$ and
$\theta_{k,l}$ are referred to as phases and angles of
rotations, respectively. This way, the complex precoding matrix
$\vec{V}_t$ can be represented compactly using $m^2$
real scalar parameters $(\phi_{k,l},\theta_{k,l})$, that can be tracked
more effectively using \eqref{eq1:syspreadapt}. However, these
parameters are still frequency dependent, and the tracking has to be
performed for all or pilot subcarriers. Moreover,
directly interpolating the frequency domain parameters may not result in
accurate representation of the precoding filter characteristics.

Another estimation approach that is suited for 5G systems employs the
angle-delay based channel estimation
approach~\cite{mo2017channel}. This approach exploits the sparse
nature of the 5G channel in the discrete-time domain to estimate the
channel coefficients. Since this method involves estimation of the complete
channel, we employ this to find only the precoder while adaptively tracking the temporal
variation of the channel coefficients.
Because of the restrictions on the amount of information for the feedback, adaptive
quantisation has to be applied. A major challenge of using
this approach is that it approximates the channel
in the discrete-time domain to only a few initial coefficients (channel taps).
This, along with adaptive quantisation results in an inaccurate precoder representation,
when reconstructing the precoder for all subcarriers at the transmitter end.

In this work, we take the approach of explicitly constructing a matrix
all-pass filter and track that filter instead, and we show that the
proposed approach requires a smaller number of parameters to feed back
while obtaining achievable rates in excess of those achieved using the
geodesic~\cite{khaled2005interpolation}, Givens rotations~\cite{madan2020scalar,roh2007efficient},
and angle-delay based method~\cite{mo2017channel}.

\section{Matrix Lattice filter}
\label{sec:matrixfilter}
\subsection{Time-domain Realisation using SNIP}
\label{subsec:lccde}
As discussed in the previous section, there are several issues when
using purely frequency domain based interpolation and design
techniques for matrix all-pass filters in general. This is
particularly applicable to situations wherein an accurate time-domain
characterisation of the filter may exist, as is often the case with 5G
mmW systems that frequently exhibit a predominant line-of-sight
characteristic. In these situations, the time domain characterisation is
not only more compact, but can also capture the response of the
precoder (when viewed as a matrix filter) more accurately. To this
end, we adapt the approach described in~\cite{unpublishedkeyS} to
design the precoder as a time-domain filter, as discussed below.

\begin{theorem}
\label{thm:thm1}
Given unitary matrices $\vec{V}(\e{j\omega_k})$ and
group-delay matrices $\vec{F}(\e{j\omega_k})$, $~k \in \{0,~\dots,M-1\}$,
there exist polynomials $\vec{N}(z)$ and
$\vec{D}(z)$ such that
$\vec{V}(\e{j\omega_k}) = \vec{N}(\e{j\omega_k})
  \vec{D}^{-1}(\e{j\omega_k})$, where $\vec{V}$ is obtained
from SVD of channel matrix
$\vec{H}$. This is called subspace Nevanlinna-Pick
interpolation (SNIP) approach to factorise unitary matrix
$\vec{V}(\e{j\omega})$ into rational matrix polynomials
$\vec{N}(\e{j\omega})$ and
$\vec{D}(\e{j\omega})$, where $\vec{N}$ and $\vec{D}$ satisfy the following
condition:
{
\begin{align*}
    \vec{V}^H(\e{j\omega}) \vec{V}(\e{j\omega}) &= \vec{I}_m \\
    \vec{N}^H(\e{j\omega})\vec{N}(\e{j\omega}) - \vec{D}^H(\e{j\omega})\vec{D}(\e{j\omega}) &= \vec{0}_m, \forall \omega \in [-\pi,~\pi).
  \end{align*}}
\end{theorem}
\begin{proof}This has been proved in \cite{unpublishedkeyS}.\end{proof}
\begin{figure}[hbt]
    \centering
    \includegraphics[width=\imgwidth]{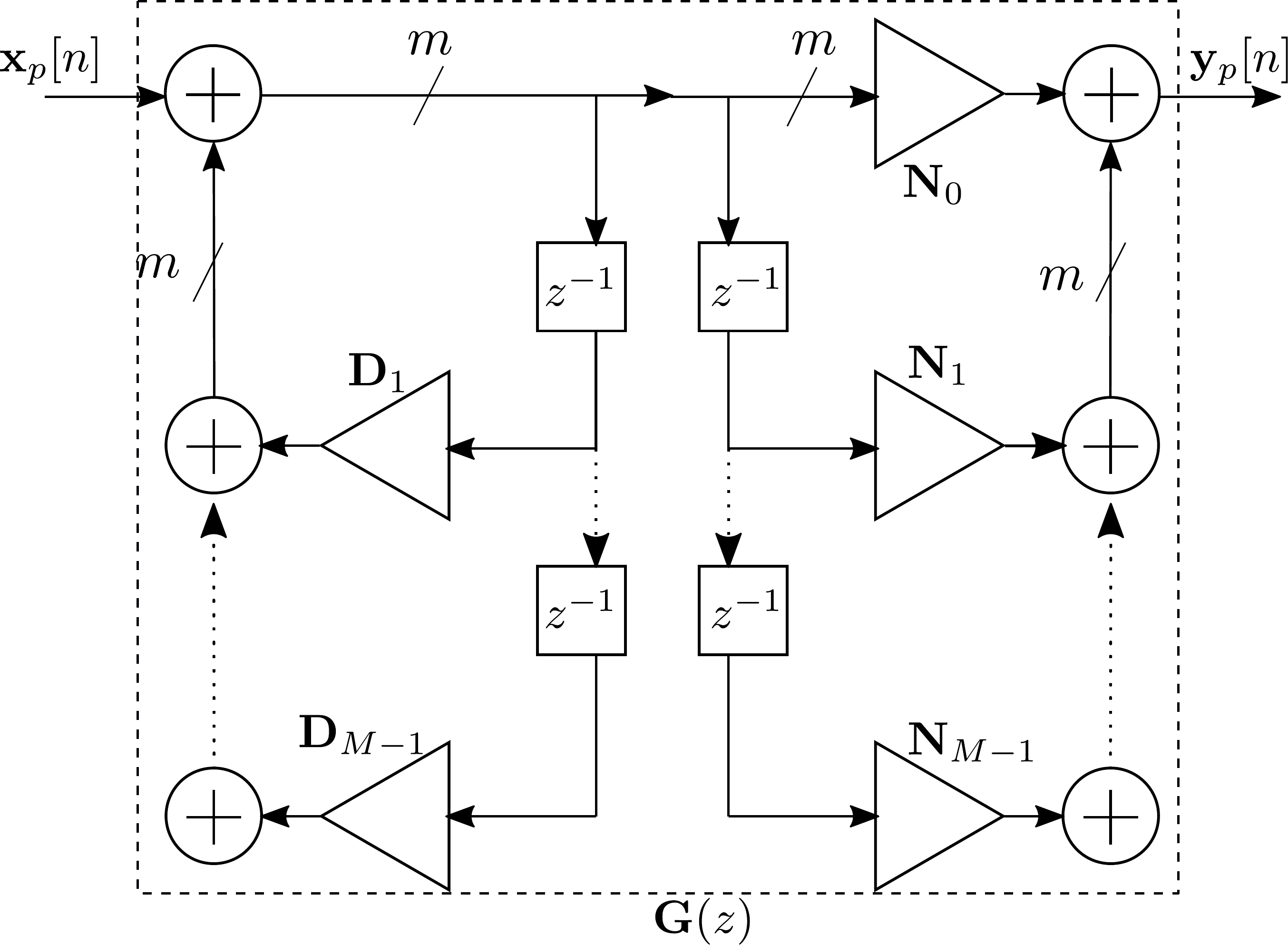}
    \caption{LCCDE (Direct Form II) implementation of matrix filter with
    right co-prime factorisation ($\vec{V}(z) = \vec{N}(z) \vec{D}^{-1}(z)$).}
    \label{fig:matrix_lccde_block}
\end{figure}

The SNIP approach from Theorem~\ref{thm:thm1}, ensures that we obtain a
\emph{realisable} filter that captures the precoder characteristics, in
the sense that $\vec{N}(\e{j\omega})$ and $\vec{D}(\e{j\omega})$ are
matrix polynomials that can be used to obtain a LCCDE that precisely
satisfies the filter constraints at $\omega_0,~\dots,~\omega_{N-1}$ while
also guaranteeing that $\vec{N}(\e{j\omega})\vec{D}^{-1}(\e{j\omega})$ is
unitary for all $\omega ~\in~[-\pi,~\pi)$. Therefore, if the precoder matrices'
frequency domain response can be characterised as a realisable matrix all-pass
filter, the SNIP approach is likely to capture this more accurately than a
frequency domain interpolation~\cite{unpublishedkeyS}. An additional
benefit is that, if the SNIP is performed using only $M$ points, the
resulting matrix all-pass filter can be realised using $M$ poles and
zeros, that can also be implemented in the time-domain using the Direct
Form II as opposed to a frequency domain approach that employs DFT-IDFT
operations.

\begin{figure}[hbt]
    \centering
    \includegraphics[width=\imgwidth]{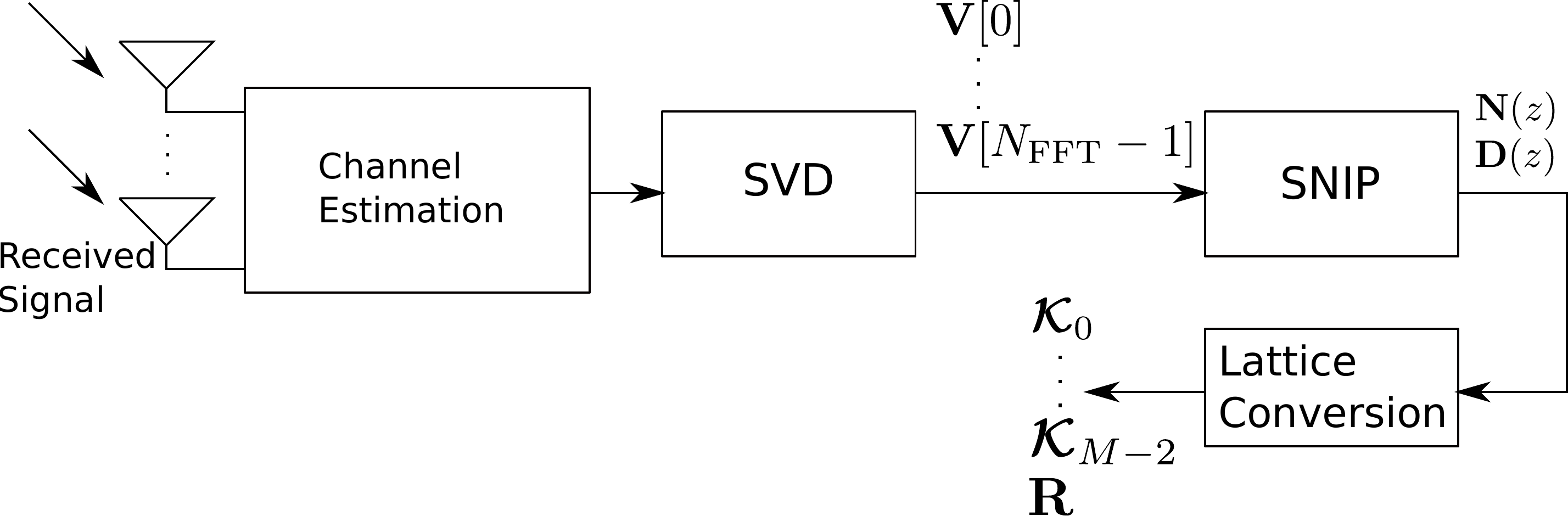}
    \caption{Process of parametrising the precoding matrices $\vec{V}[0],\ldots,\vec{V}[N_{\text{FFT}}-1]$ using matrix
    lattice parameters $\vecsym{\K}_0,\ldots,\vecsym{\K}_{M-2}, \vec{R}$ where $M \ll N_{\text{FFT}}$}
    \label{fig:lattice_block}
\end{figure}

We denote the all-pass filter as $\vec{G}(\e{j\omega})$, which is the
precoding matrix $\vec{V}(\e{j\omega})$. $\vec{x}_p$ and $\vec{y}_p$ are
respectively the input and output of the precoder $\vec{G}$.
\figurename~\ref{fig:matrix_lccde_block} depicts this filter with Direct
Form II realisation. Here,
\begin{equation}
    \vec{N}(\e{j\omega}) = \sum\limits_{i=0}^{M-1} \vec{N}_i \e{-ij\omega},~\vec{D}(\e{-j\omega}) = \sum\limits_{i=0}^{M-1} \vec{D}_i \e{-ij\omega},
    \label{eq1:matrixfilterlccde}
  \end{equation}
where $\vec{N}_k, \vec{D}_k, k=0,1,\dots,M-1$ are
matrices, and $\vec{D}_0 = \vec{I}_m$.
Although the Direct Form II is an effective way of realising a discrete-time filter, there
exist certain limitations in this representation for our application,
as listed below.
\begin{enumerate}
\item When transforming the filter representation from frequency-domain
to time-domain realisation, the stability of
  the filter becomes important, even for an all-pass filter. For
  scalar all-pass filters, an explicit relationship exists between the
  numerator and denominator polynomials (as discussed in
  Section~\ref{subsec:adv}) that ensures the filter stability
  even after quantisation. Unfortunately, for the matrix all-pass
  filter case, there is no such direct relation between $\vec{N}(\e{j\omega})$
  and $\vec{D}(\e{j\omega})$ polynomials. Hence, it becomes
  difficult to comment on the stability directly from these
  coefficients.
\item Because of the absence of a ``convenient'' structure for $\vec{N}(\e{j\omega})$
and $\vec{D}(\e{j\omega})$, directly tracking the filter coefficients
while preserving the unitary nature of $\vec{G}(\e{j\omega})$ and its
stability are also difficult.
\end{enumerate}
To address these limitations, we propose the use of lattice structures
to represent $\vec{G}(\e{j\omega})$ instead of the Direct Form II
realisation. The process of converting the precoding matrices $\vec{V}[0],\ldots,\vec{V}[N_{\text{FFT}}-1]$
to matrix lattice parameters $\vecsym{\K}_0,\ldots,\vecsym{\K}_{M-2},\vec{R}$ is shown
in \figurename{~\ref{fig:lattice_block}}. The advantage of doing so is that
the number of lattice parameters are much smaller than the number of subcarriers,
resulting in lesser feedback requirement. Moreover, the lattice parameters are
suitable for adaptive quantisation, without degrading the unitary nature of
the overall response. These claims are explained in Section~\ref{subsec:adv}.

For simplicity of derivation, we consider all inputs and
outputs are of same dimensions $m$, which implies we will only
deal with $m \times m$ square matrices, although these
techniques are extensible to the case of non-square precoders as well
using approaches similar to those discussed in~\cite{roh2007efficient}.
We will also use
$z$ instead of $\e{j\omega}$ to represent all the
matrix polynomials in terms of the general complex argument $z$.

\subsection{Lattice Structure Realisation}
\label{subsec:lattice}
To understand the lattice realisation for the matrix all-pass filter, we
first look into the scalar case. The transfer function of a stable scalar
all-pass filter can be given as
\begin{equation}
    H_d(z) = \frac{N_d(z)}{D_d(z)} = \frac{z^{-d} D_d^*(1/z^*)}{D_d(z)}
    \label{eq6:matrixfilterlattice}
\end{equation}
where $d$ is the polynomial degree and $(\cdot)^*$
denotes the complex conjugate~\cite{regalia1988digital}. Since the filter
is stable, the roots of $D_d(z)$ (poles of $H_d(z)$) lie inside the unit-circle
in the complex $z$-plane. One can synthesise the lattice structure to realise
the given filter of degree $d$ iteratively, as shown below.
\begin{equation*}
    z^{-1} H_{k-1}(z) = \frac{H_k(z) - C_{k-d}}{1 - C_{k-d} H_k(z)}, ~ k \in \{d,~d-1,~\dots,~1\}
  \end{equation*}
where $C_{k-d} = H_k(\infty)$ are the lattice
parameters. Since the poles of $H_d(z)$ lie inside the
unit-circle, it can be verified that $H_{d-1}(z)$ is also
stable, and $|C_k| < 1$~\cite{regalia1988digital}. Therefore, inferring
the stability of
the filter using the Direct Form implementation (in terms of
$N(z)$ and $D(z)$) requires computation of roots of
a $d$-degree polynomial, whereas for the lattice realisation, only
verifying that $|C_k| < 1$ is enough. Moreover, it is
convenient to adapt and quantise the lattice parameters
$C_k$, since it combines the effect of adapting and
quantising both $N(z)$ and $D(z)$, while ensuring
the stability and all-pass nature. These are some of the key
advantages of lattice based filter realisations, and these can be
adapted effectively to the matrix case to represent the precoder
variations across time and frequency efficiently.

\begin{figure}[hbt]
    \centering
    \includegraphics[width=\imgwidth]{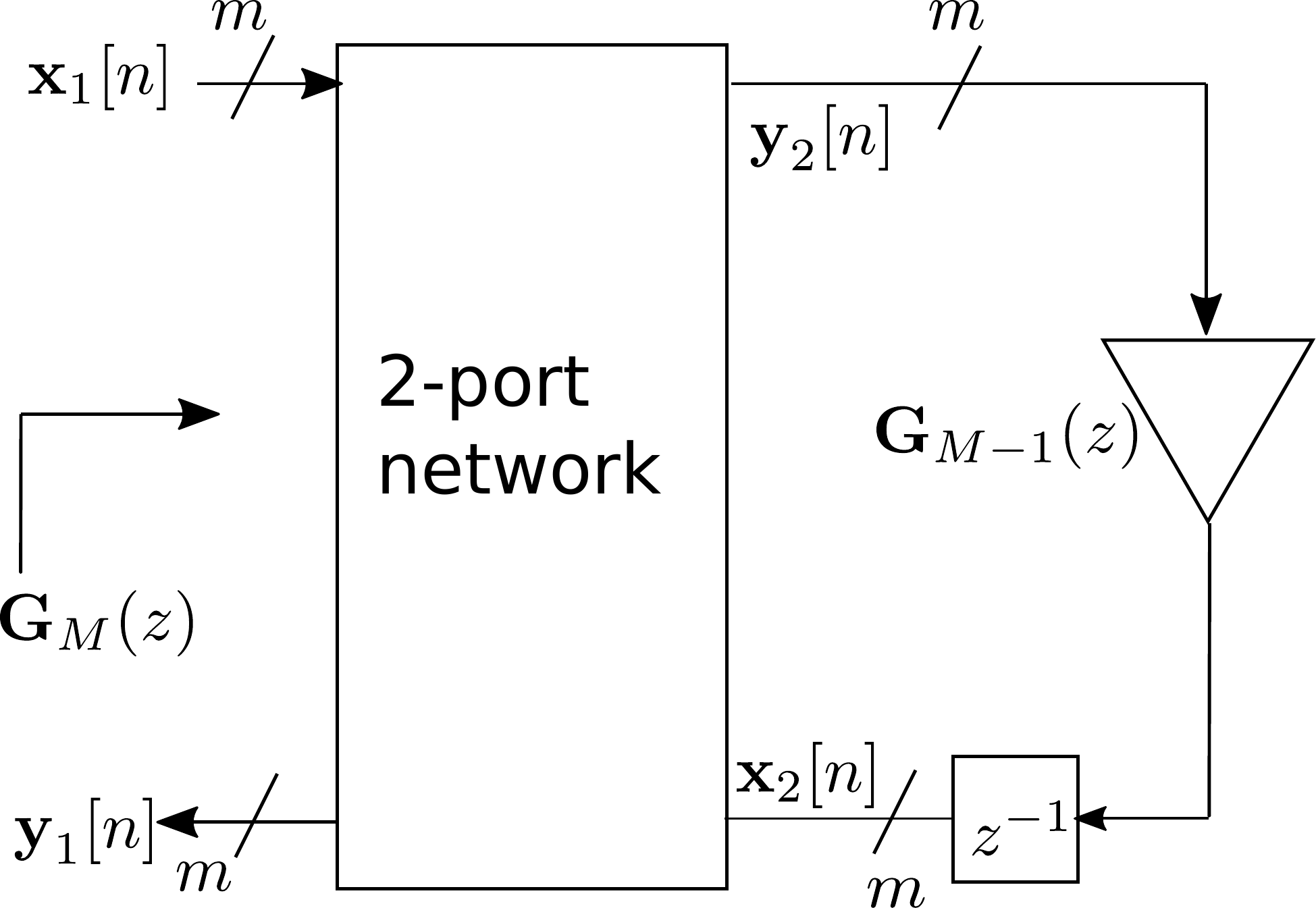}
    \caption{2-port representation of lattice filter structure.}
    \label{fig:2-port}
\end{figure}

The matrix lattice structure can be constructed from the Direct
Form II by considering a 2-port system, as shown in
\figurename~\ref{fig:2-port}, where the output at port-2
($\vec{y}_2$) is connected to input at port-2
($\vec{x}_2$). The transfer function at port-2 is considered
to be of length $M-1$ and the overall transfer function is of
length $M$.

\begin{figure}[hbt]
    \centering
    \includegraphics[width=\imgwidth]{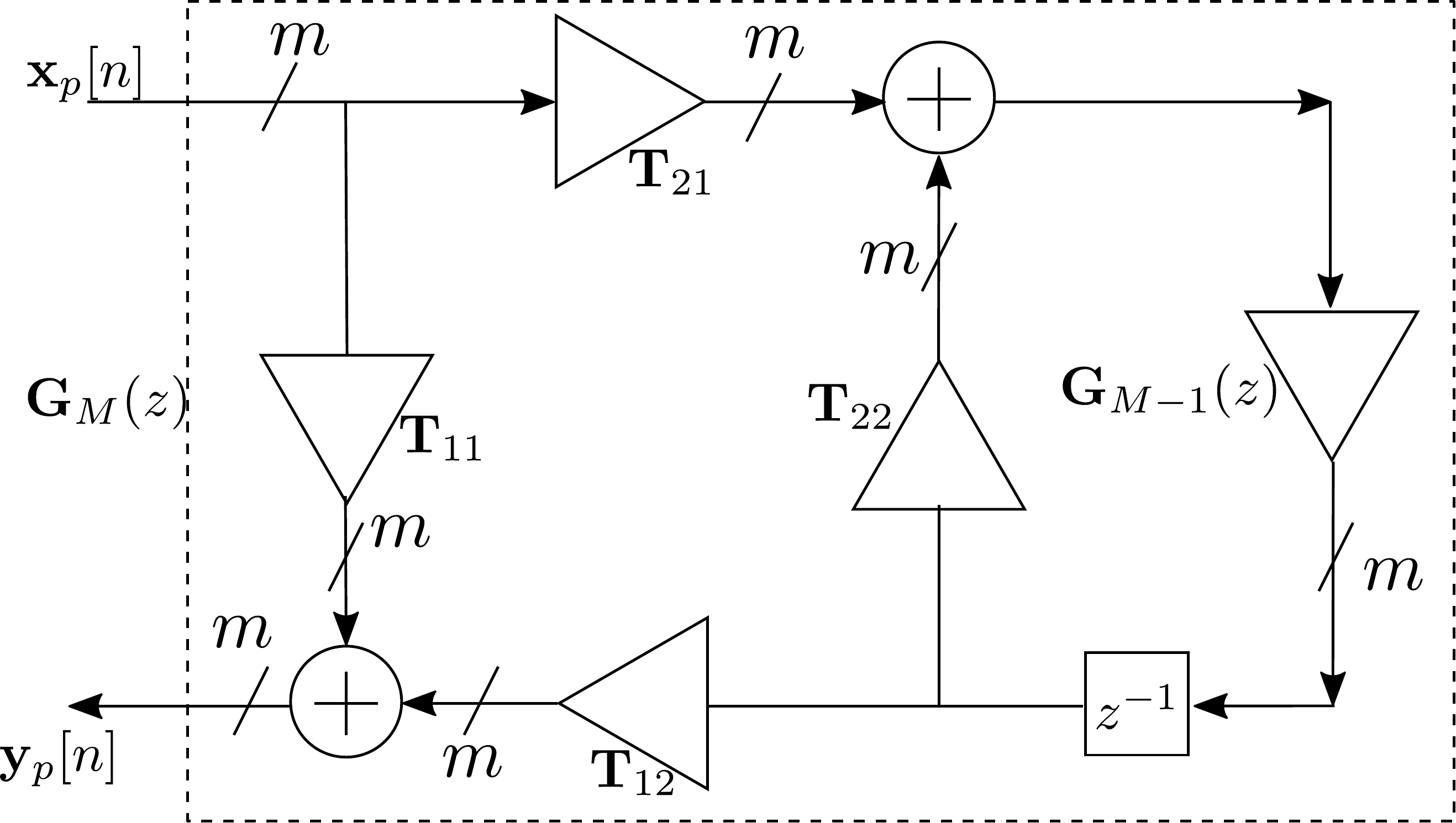}
    \caption{Realising a single stage of matrix lattice filter using
    T-parameters.}
    \label{fig:lattice_T}
\end{figure}

The 2-port system can be realised using the transmission
parameters (T-parameters) as shown in
\figurename~\ref{fig:lattice_T}. Denoting the input and output vectors
as $\vec{x}_k(z)$ and $\vec{y}_k(z)$ for
$k = 1,2$ respectively, we can write
\begin{equation}
    \vec{y}[n] = \begin{bmatrix} \vec{y}_1[n] \\ \vec{y}_2[n] \end{bmatrix} = \begin{bmatrix} \vec{T}_{11} & \vec{T}_{12} \\ \vec{T}_{21} & \vec{T}_{22} \end{bmatrix} \begin{bmatrix} \vec{x}_1[n] \\ \vec{x}_2[n] \end{bmatrix} = \vecsym{\T} \vec{x}[n]
    \label{eq1:matrixfilterlattice}
\end{equation}
where $\vec{x}_1[n] = \vec{x}_p[n]$ and $\vec{y}_1[n] = \vec{y}_p[n]$.
Similar to scalar lattice filter realisations, at the $k$th stage the
T-parameters are related to each other and can be characterised by a
common matrix $\K_k$ as shown in \eqref{eq5:matrixfilterlattice}~\cite{vaidyanathan1985general}.
\begin{equation}
    \T_k = \begin{bmatrix} \K_k & (\vec{I}_m - \K_k\K_k^H)^{\frac{1}{2}} \\ (\vec{I}_m - \K_k^H \K_k)^{\frac{H}{2}} & -\K_k^H
    \end{bmatrix}
    \label{eq5:matrixfilterlattice}
\end{equation}
where $(\cdot)^{\frac{1}{2}}$ denotes the matrix square root of a
complex matrix, $(\cdot)^{\frac{H}{2}}$ and $(\cdot)^{-\frac{H}{2}}$
denotes square root and inverse of the square root of the Hermitian transpose
of the matrix, respectively; $\K_k$ is of size $m\times m$ and $\T_k$
is of size $2m\times 2m$.

As discussed earlier, several limitations exist when implementing the
precoding using either the frequency-domain methods discussed in
Section~\ref{sec:sys} or the Direct Form II implementation listed in
Section~\ref{subsec:lccde}. This is particularly affects the
situations where adaptation of filter coefficients is performed, since
ensuring that the resulting filter is stable requires numerical
verification after every adaptation. To address this and other issues,
we propose the lattice based realisation, since it addresses the
issues with past approaches, as discussed in Section~\ref{subsec:adv}.

\begin{figure}[hbt]
    \centering
    \includegraphics[width=0.9\columnwidth]{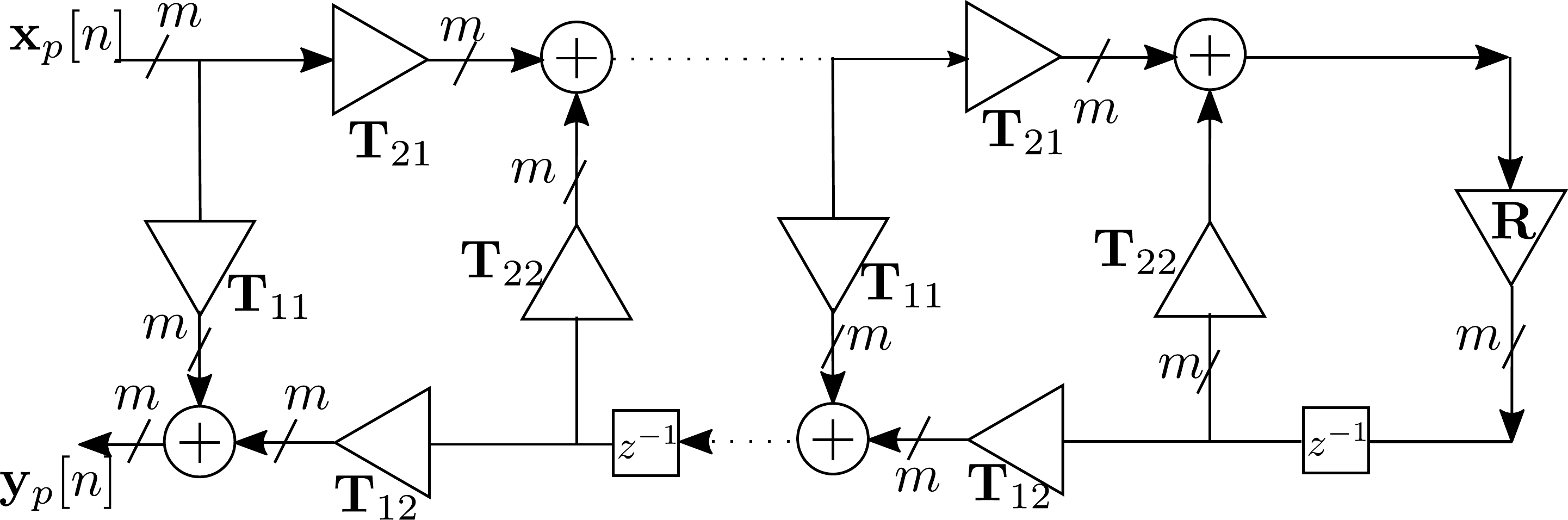}
    \caption{Overall matrix lattice filter of length $M$.}
    \label{fig:latticeF}
\end{figure}

We next discuss the approach to obtain the T-parameters from
$\vec{N}(z)$ and
$\vec{D}(z)$. Algorithm~\ref{alg:latticesteps} captures the
step-wise process to iteratively obtain the T-parameters.
\begin{algorithm}
    \caption{Iteratively compute the matrix lattice parameters}
    {\begin{algorithmic}
        \State Start with unitary $\vec{G}_n(z) = \vec{N}_n(z) \vec{D}_n(z)^{-1}$.
        \Repeat
        \State Multiply $\vec{N}_n(z)$ and $\vec{D}_n(z)$ by $\vec{D}_0^{-1}$ such that $\vec{D}_0 = \vec{D}_n(\infty) = \vec{I}_m$.
        \State $\K_{n-k} = \vec{G}_k(\infty) = \vec{N}_k(\infty) = \vec{N}_0$
        \State Find T-parameters using \eqref{eq5:matrixfilterlattice}.
        \State Find $\vec{\hat{D}}_n = \left( \vec{T}_{21}^H \vec{T}_{21}\right)^{-1} \left( \vec{D}_n(z) - \K_{n-k}^H \vec{N}_n(z) \right)$.
        \State Find $\vec{\hat{N}}_n = \vec{N}_n(z) - \K_{n-k} \vec{\hat{D}}_n$.
        \State $\vec{D}_{n-1}(z) = \vec{T}_{21} \vec{\hat{D}}_n$
        \State Remove the last matrix, which will be $\vec{0}_m$.
        \State $\vec{N}_{n-1}(z) = \left( \vec{T}_{12} + \K_{n-k}\vec{T}_{21}^{-1}\K_{n-k}^H \right)^{-1} \vec{\hat{N}}_n$
        \State Remove the first matrix which will be $\vec{0}_m$.
        \State Find $\vec{G}_{n-1}(z) = \vec{N}_{n-1}(z) \vec{D}_{n-1}(z)^{-1}$.
        \Until $\vec{G}_{0}(z)$ is reached.
        \State $\vec{R} = \vec{G}_0(z)$
    \end{algorithmic}}
    \label{alg:latticesteps}
\end{algorithm}
The algorithm largely corresponds to the traditional approach of
converting Direct Form II realisations to lattice structures for scalar
filters, with minor differences arising due to the matrix case. We
begin with the SNIP based filter that is obtained using
Theorem~\ref{thm:thm1}, referred to as $\vec{G}_n(z)$, that yields
$\vec{N}_n(z)$ and $\vec{D}_n(z)$. In each iteration, the algorithm
computes one of the $\K_k$ parameters, and ``removes'' the
contribution of this parameter from the original filter to obtain the
all-pass filter $\vec{G}_{n-1}(z)$, that has order one less than
$\vec{G}_n(z)$.  After $M-1$ stages, the last remaining
polynomial is the residue transfer function
($\vec{R}$). Hence, the whole matrix filter can be
characterised by $M$ parameters (viz. $\K_0$,
  $\K_1$, \ldots $\K_{M-2}$ and one
$\vec{R}$), as opposed to LCCDE implementation which requires
$2M$ parameters ($M~\vec{N}_k$ and
$M ~\vec{D}_k$). The overall lattice filter is shown in
\figurename~\ref{fig:latticeF}.

To preserve the unitary nature for each lattice stage, the following conditions should be satisfied:
\begin{subequations}
\begin{align}
    \vec{T}_{11}^H \vec{T}_{11} + \vec{T}_{21}^H \vec{T}_{21} &= \vec{I}_m \label{eq3a:matrixfilterlattice} \\
    \vec{T}_{12}^H \vec{T}_{12} + \vec{T}_{22}^H \vec{T}_{22} &= \vec{I}_m \label{eq3b:matrixfilterlattice} \\
    \vec{T}_{11}^H \vec{T}_{12} + \vec{T}_{21}^H \vec{T}_{22} &= \vec{0}_m \label{eq3c:matrixfilterlattice}.
\end{align}
\label{eq3:matrixfilterlattice}
\end{subequations}
The proof and conditions for real T-parameters can be adapted from
the results in~\cite{vaidyanathan1985general} and can be suitably modified
for complex T-parameters. The conditions in \eqref{eq3:matrixfilterlattice} can
be combined into a single condition as
\begin{equation}
    \T^H \T = \begin{bmatrix} \vec{I}_m & \vec{0}_m \\ \vec{0}_m & \vec{I}_m\end{bmatrix} = \vec{I}_{2m}.
    \label{eq4:matrixfilterlattice}
\end{equation}

\subsection{Advantages of Lattice Realisation}
\label{subsec:adv}
We now discuss how the use of lattice realisations addresses the
issues of ensuring stability, adaptability, and easy quantisation of
the precoding all-pass filter when compared to the past approaches.

\subsubsection{Stability}
Even though precoders are implemented as all-pass filters, ensuring
filter stability is essential, since a practical implementation that
does not ensure stability could result in saturation and loss in
accuracy that could compromise performance significantly. This is even
more important in the case where these filters are parameterised and
the parameters are tracked, since some parameter updates could cause
loss in stability.  Verifying the stability of causal filters in
general requires computation of roots of the denominator polynomial to
ensure that the poles are within the unit circle. This is especially
applicable to matrix all-pass filters, wherein it is difficult to
infer the stability directly from the filter coefficients $\vec{N}(z)$
and $\vec{D}(z)$. Explicitly, such an approach would require
evaluating all roots of $\det{(\vec{D}(z))}$ and ensuring that they
are all within the unit circle, which is computationally intensive. In
fact, root finding algorithms for the most general case requires
either computation of solvents, which may not be guaranteed to exist
in all cases~\cite{dennis1978algorithms}. Numerical approaches such as
Newton's methods can be faster~\cite{kratz1987numerical}, but may
suffer from convergence related issues, especially when solving equations
in higher dimensions, such as the ones that we consider here. In contrast, by
realising and representing the filter in the lattice form, the lattice
parameters $\K_k$ can directly capture the stability information, much
like in the case of scalar lattice structures. The scalar filter is
considered to be stable if the lattice parameters for all stages
$C_k$, satisfy $|C_k| < 1$. Similarly, for the matrix filter case, we
have~\cite{vaidyanathan1985general}:
\begin{equation}
    \K_k^H \K_k \prec \vec{I} \text{ and } \K_k \K_k^H \prec \vec{I}
    \label{eq1:matrixfilteradvstable}
\end{equation}
which implies that $\vec{I} - \K_k^H\K_k$ and
$\vec{I} - \K_k\K_k^H$ are positive definite
matrices. We remark here that we use non-degenerate version of the
    inequalities mentioned in~\cite{vaidyanathan1985general}. This
    condition also ensures the matrix roots in
\eqref{eq5:matrixfilterlattice} exist. In other words, all the
singular values of $\vec{T}_{ij}$ lie in $[0, 1]$, for all
stages. Thus, the stability can be verified by ensuring that
\eqref{eq5:matrixfilterlattice} without having to perform any
additional numerical computations.

\subsubsection{Quantisation}
Another issue with both the frequency based approaches as well as the
Direct Form II implementations is that the parameters are not easy to
quantise. In the frequency based approach, not capturing the
time-domain structure would imply that the precoder should be
quantised in the frequency domain for more subcarriers to ensure
accuracy of reconstruction, thereby resulting in a much larger
feedback burden. In the Direct Form II realisation that can be
inferred in a straightforward manner by applying
Theorem~\ref{thm:thm1}, we note that a useful structure that relates
$\vec{N}(z)$ and $\vec{D}(z)$ and is useful for quantisation is
absent. This is in contrast to the scalar all-pass filter case,
wherein the numerator and denominator polynomials $N(z)$ and $D(z)$
are in reverse order and conjugate~\cite{regalia1988digital} as shown
in \eqref{eq6:matrixfilterlattice}.  Hence, if we were to quantise one
set of coefficients, say $N(z)$, it directly captures $D(z)$
implicitly, owing to the fact that the absolute value of the filter
response on the unit circle is unity. Such a direct constraint is not
present in the matrix analogues, viz. $\vec{N}(z)$ and $\vec{D}(z)$.
In addition, directly quantising the coefficients of $\vec{N}(z)$ and
$\vec{D}(z)$ could result in a filter that may not be all-pass for all
$z = e^{j\omega}$, since the conditions in Theorem~\ref{thm:thm1} have
to be valid for all $\omega \in [-\pi, \pi)$. When considering the
lattice based approach, all of these issues can be addressed easily.
Since each lattice stage is characterised by only one parameter, viz.
$\K_k$, the information fed back from the receiver to the transmitter
can be reduced, as we show in Section~\ref{sec:results}. Moreover,
since $\K_k$ completely parameterises the precoding all-pass filter,
the need to track $\vec{N}(z)$ and $\vec{D}(z)$ separately is
obviated, since $\K_k$ can be tracked adaptively in order to capture
the filter characteristics fully. In addition, the lattice structure
guarantees that the resulting filter will be all-pass even when $\K_k$
are quantised.

To quantise and adapt the $\K_k$, we use the parameter update approach
described in \cite{roh2007efficient}. For a slowly varying channel
$\vec{H}_t$, the parameters $\K_k$ also vary slowly, as shown in
\eqref{eq3:sysmodel} according to an AR(1) process. Similar to that in
\eqref{eq1:syspreadapt}, we can write the adaptive quantisation of
$\K_k$ for the $k$th stage as
\begin{equation}
	\begin{split}
    \hat{\K}_{k_t} &= \hat{\K}_{k_{t-1}} + \vec{\Delta}_{k_t} \vec{\Gamma}_{k_t},\\
    ~&\text{where}~\vec{\Gamma}_{k_t} = \begin{cases} \sigma\vec{\Gamma}_{k_{t-1}}&: \vec{\Delta}_{k_t} = \vec{\Delta}_{k_{t-1}} \\ \vec{\Gamma}_{k_{t-1}}/\sigma&: \vec{\Delta}_{k_t} \neq \vec{\Delta}_{k_{t-1}}\end{cases}
	\end{split}
    \label{eq1:matrixfilteradvquant}
\end{equation}
where $\vec{\Delta}$ and $\vec{\Gamma}$ are complex matrix updates,
that correspond to $\beta$ and $\mu$
in \eqref{eq1:syspreadapt}. The comparisons ($=$ and $\neq$) and
multiplication ($\vec{\Delta}_{k_t} \vec{\Gamma}_{k_t}$) are element-wise
and independent for real and imaginary parts of the complex matrices.
$\sigma$ is chosen such that the entries of $\K_k$ do not exceed $1$,
and \eqref{eq1:matrixfilteradvstable} is satisfied. This ensures that
the unitary nature and stability of the lattice realisation is preserved,
even after adaptive quantisation.

\subsubsection{Adaptability}
As shown in \eqref{eq3:sysmodel}, a slowly varying channel
$\vec{H}_t[n]$ can be modelled as an AR(1) process. For such channels,
it is desired to track the precoding parameters accordingly. This is
computationally costly for the previous frequency-domain based
methods, where the precoding matrix for every subcarrier has to be
tracked, or at least tracked for some subcarriers and interpolated for
the rest, thereby affecting accuracy. The time-domain realisation
helps, as the number of parameters to track ($\vec{N}(z)$ and
$\vec{D}(z)$) could be less than that for the frequency-domain
approach. However, using the lattice structure, the number of
parameters can be reduced further (from $2M$ to $M$), which can
potentially make the process faster. Moreover, the
\emph{phase-unwrapping} because of $2\pi$ jumps
required~\cite{madan2020scalar} in the frequency-domain method can
also be avoided. The adaptive quantisation thus explained in the
previous section can quantise and adapt to the slowly varying channel
by tracking the lattice parameters $\K$.

As per \eqref{eq3:sysmodel} and \eqref{eq4:sysmodel}, the rate of change is
dependent on the relative speed between the transmitter and receiver. For
larger speeds, the value of $\alpha$ is low, which means the correlation
between $\vec{H}_t[n]$ and $\vec{H}_{t-1}[n]$ drops, and the tracking parameters
$\vecsym{\Gamma}$ and $\sigma$ in \eqref{eq1:matrixfilteradvquant} changes
accordingly. For typical slow variations in channel, $\alpha \geq 0.9$.

\section{Simulation Results And Discussion}
\label{sec:results}
In this section, we present simulation results that quantify the
benefit of using the proposed adaptive precoder feedback approach, and
compare this with other adaptive precoder feedback methods that are
primarily frequency-domain based. We also verify the advantages of the
lattice representation of precoders and compare them with the
frequency based geodesic~\cite{khaled2005interpolation} and Givens
rotation~\cite{roh2007efficient,madan2020scalar} methods.

To model the 5G mmW channel characteristics, we consider the empirical
model described in~\cite{samimi20163} that specifies the power delay
profiles for a typical 28 GHz 3D channel model that was verified
experimentally. Adapting the parameters described therein
appropriately to our scenario, the resulting power-delay profile is
shown in \tablename~\ref{tab:channelprof}. The powers are normalised
with respect to the first temporal component.
\begin{table}[hbt]
\caption{Power-delay profile of the 5G millimeter-wave channel\label{tab:channelprof}}
\centering
\begin{tabular}{|c||ccccc|}
\hline
{\bf Normalized Power (\si{\deci\bel})} & 0 & -112 & -132 & -142 & -153\\
{\bf Delay (\si{\nano\second})} & 0 & 381 & 407 & 1433 & 1500 \\
\hline
\end{tabular}
\end{table}
This is a wideband
channel, and the LOS component is dominant while the reflections are
very small. For such a channel, a typical approach to precoder design
is sub-band-wise~\cite{gao2016energy}. Due to the wideband nature of
the channel, the number of OFDM subcarriers is very large so as to
ensure that the channel can be considered to be flat fading within a
subcarrier. We consider $4 \times 4$, $8 \times 8$, $12 \times 12$,
and $15 \times 15$, MIMO systems with total 4096 subcarriers. 
We consider 4 pilot subcarriers for $4 \times 4$ and $8 \times 8$ systems,
and 8 pilot subcarriers for $12 \times 12$ and $15 \times 15$
systems.
We then compare the achievable rates with precoding, with the precoder
being fed back only
for the pilot subcarriers.
In the case of the geodesic interpolation~\cite{khaled2005interpolation}
and the scalar parameterisation~\cite{roh2007efficient,madan2020scalar}
approaches, the precoder is tracked at these pilot subcarriers, and is
interpolated in the frequency domain at the
transmitter to construct the complete precoder for all
subcarriers. The pilot subcarriers are chosen equi-spaced in the
span $[0,4095]$. For the angle-delay domain precoder tracking, we first
take an inverse Fourier transform of the precoder, which gives an
equivalent discrete-time representation~\cite{mo2017channel}. We then truncate this time-domain
precoder to the number of RB considered (viz. 4 and 8). This time-domain
precoder is then quantised and tracked, and is fed back to the transmitter
to reconstruct the complete precoder for all subcarriers.

In our approach that uses the lattice structure
parameters, a time domain matrix all-pass filter is first computed at
the receiver using a different set of pilot subcarriers depending on
the filter order (equi-spaced), and then represented using lattice
structure parameters. These lattice structure parameters are
subsequently tracked by applying feedback on the $\K_k$ parameters, as
described in Section~\ref{subsec:adv}. The parameters considered
for the simulations are shown in \tablename~\ref{tab:params}.
\begin{table}[hbt]
\caption{Simulation Parameters\label{tab:params}}
\centering
\begin{tabular}{|c||c|c|c|c|}
\hline
{\bf MIMO System} & $4 \times 4$ & $8 \times 8$ & $12 \times 12$ & $15 \times 15$\\ \hline
{\bf Total subcarriers} ($N_{\text{sc}}$) & \multicolumn{4}{c|}{4096} \\ \hline
{\bf Pilot subcarriers} ($N_{\text{p}}$) & 4 & 4 & 8 & 8 \\
{\bf Total feedback bits (Geodesic)} & 128 & 512 & 2304 & 3600 \\
{\bf Total feedback bits (Givens)} & 64 & 256 & 1152 & 1800 \\ \hline
{\bf Lattice filter order} ($N_{\text{ord}}$) & 3 & 5 & 7 & 7 \\
{\bf Total feedback bits (Lattice)} & {\bf 96} & {\bf 640} & {\bf 2016} & {\bf 3150} \\ \hline
\end{tabular}
\end{table}

\begin{figure}[hbt]
    \centering
    \includegraphics[width=\imgwidth]{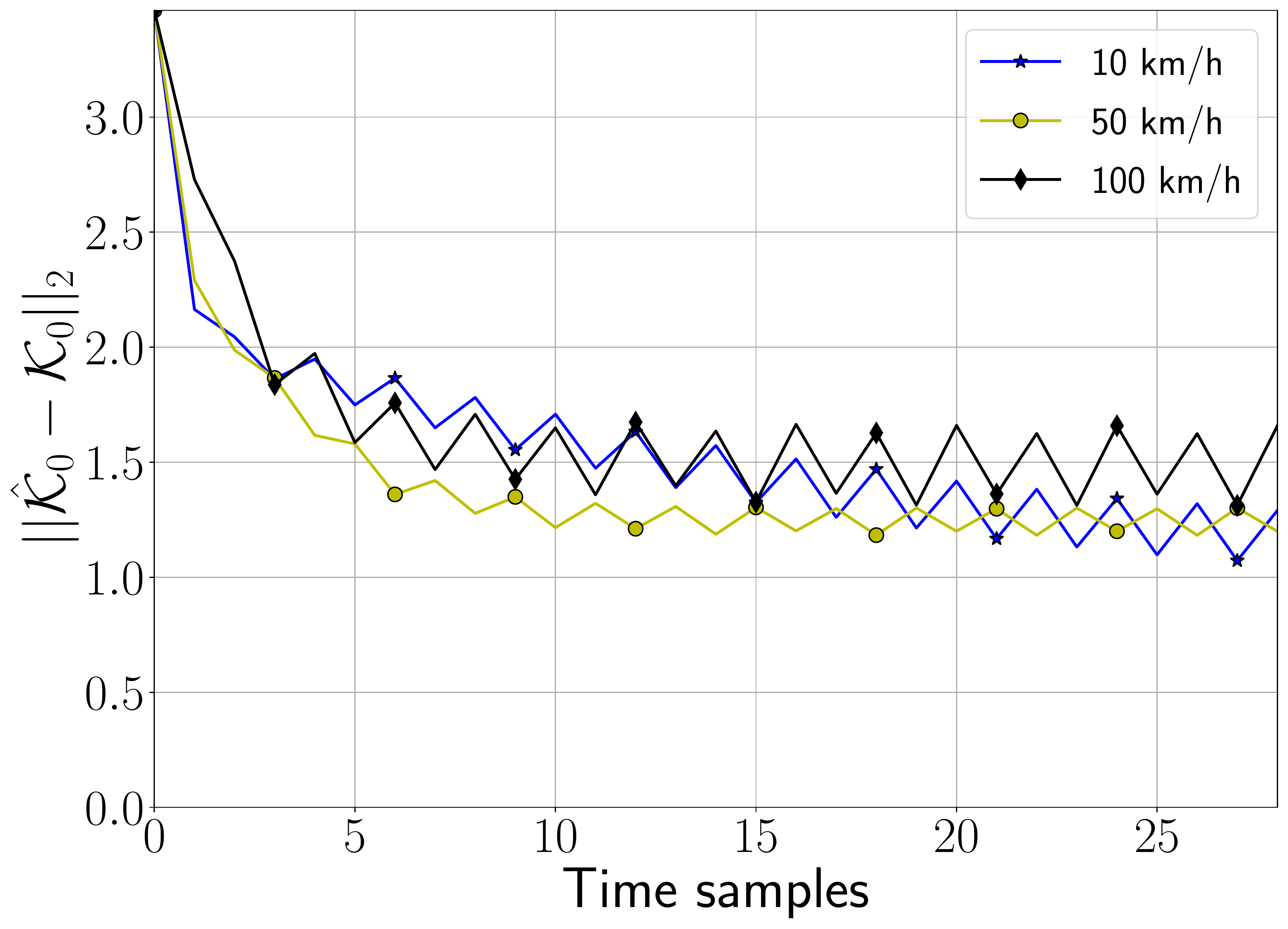}
    \caption{Error in the estimated $\hat{\K}$ for $12 \times 12$ precoder
      matrices for speeds 10, 50, and 100~\si{\kilo\meter/\hour}.}
    \label{fig:slowK}
\end{figure}

\begin{figure}[hbt]
    \centering
    \includegraphics[width=\imgwidth]{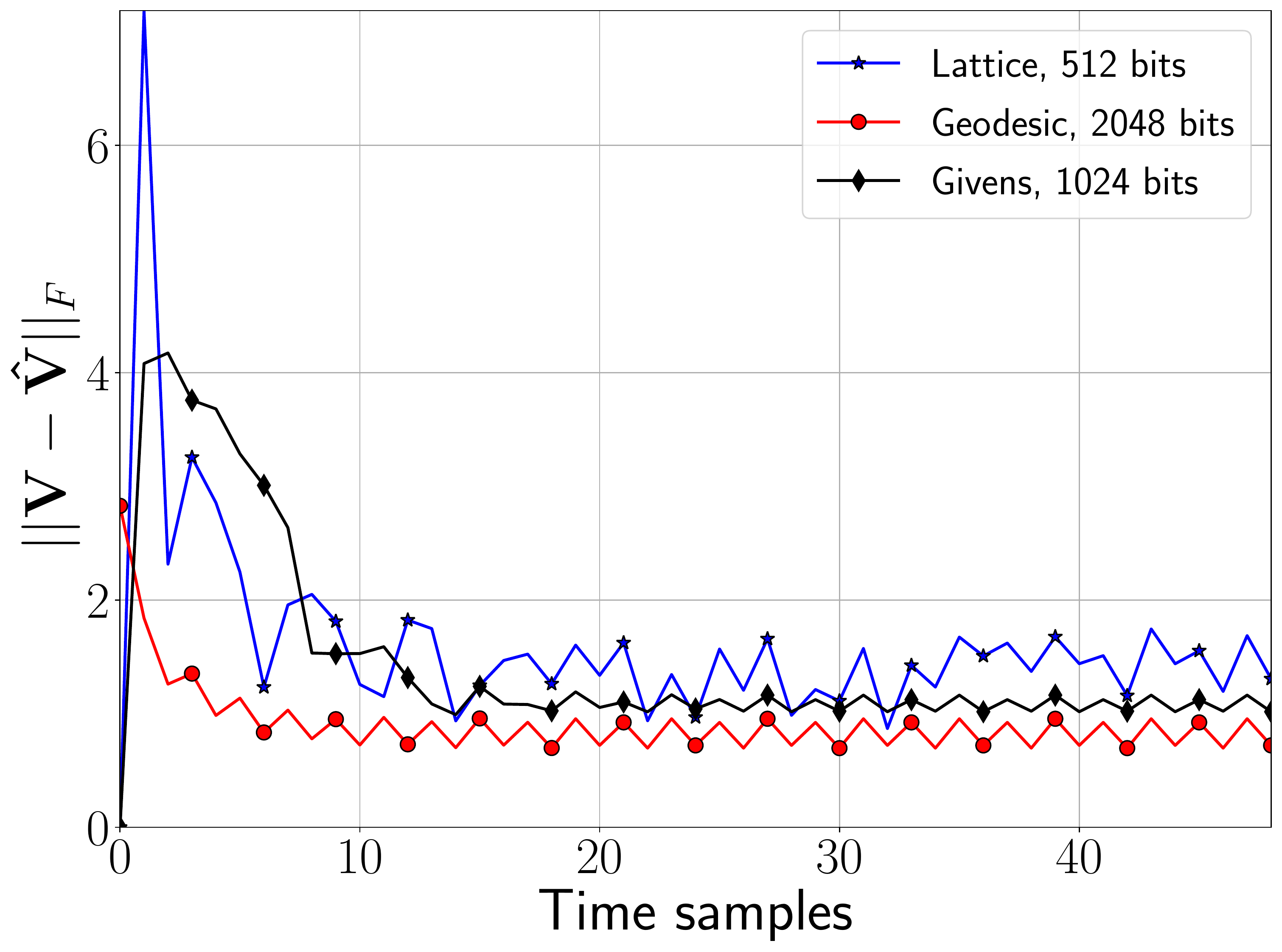}
    \caption{Comparison of the difference between the actual and estimated
    precoders using the Lattice, Geodesic, and Givens methods for an $8 \times 8$
    MIMO system for speed 50~\si{\kilo\meter/\hour}.}
    \label{fig:ncomp}
\end{figure}

\begin{figure}[hbt]
    \centering
    \includegraphics[width=\imgwidth]{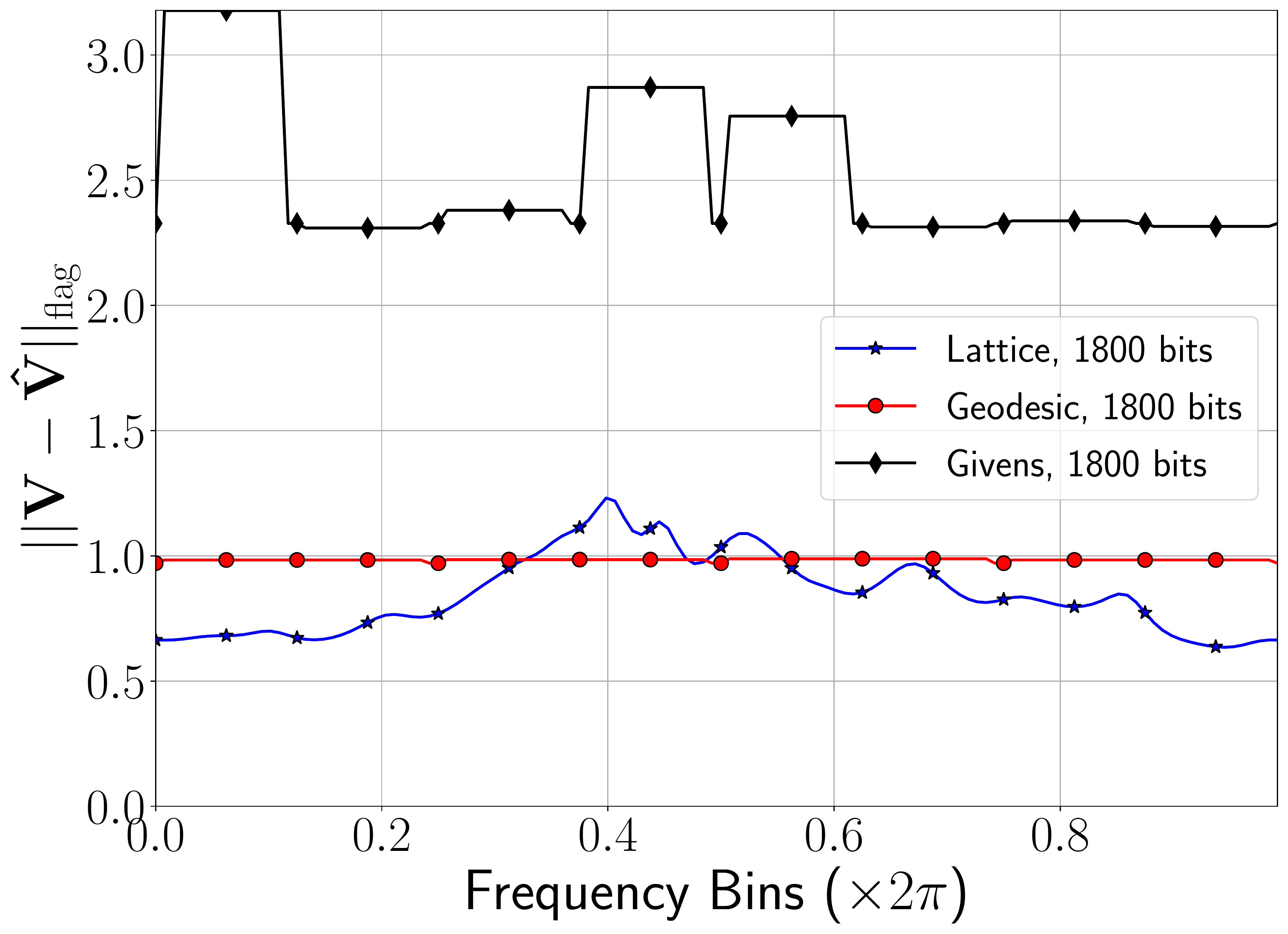}
    \caption{Comparison of the flag distance between the actual and estimated
    precoders using the Lattice, Geodesic, and Givens methods for a $15 \times 15$
    MIMO system for speed 50~\si{\kilo\meter/\hour} across all subcarriers in $[0,2\pi]$.}
    \label{fig:nflag}
\end{figure}

We first consider a slowly varying channel, as described in
\eqref{eq3:sysmodel} and \eqref{eq4:sysmodel} for typical speeds
$v \in \{ 10, 50, 100\}$~\si{\kilo\meter/\hour}, and
$T_s \approx 75$~\si{\micro\second}. We use the adaptive
quantisation as per \eqref{eq1:matrixfilteradvquant} to track the
precoder for both geodesic and lattice methods, wherein the parameter
$\sigma$ is tuned optimally, depending on the speed. As shown in
Table~\ref{tab:params}, the geodesic and Givens parameters based
approaches employ precoder interpolation from the quantised precoders
from four pilot subcarrier locations for the $4\times 4$ and
$8\times 8$ systems, and eight subcarrier locations for larger MIMO
configurations. In all the cases, we note that the lattice based
approach requires a significantly smaller number of parameters and
lower feedback, owing to the fact that they capture the precoder
characteristics in a more compact manner. This is confirmed when we
observe the convergence of the filter parameters by viewing the precoder
adaptation error, as shown in \figurename~\ref{fig:slowK}, and the estimated
precoder error, over time, as shown in \figurename~\ref{fig:ncomp}; and across
subcarriers as shown in \figurename~\ref{fig:nflag}.

To further highlight the effectiveness
of the proposed lattice based approach to track the precoder for these channels, we
have presented the performance of this approach with multiple levels of feedback, viz.
64 and 96 bits per OFDM frame in the $4 \times 4$ situation depicted in \figurename~\ref{fig:speed1_4},
384 to 708 bits per OFDM frame for the $8 \times 8$ situation in \figurename~\ref{fig:speed1_8},
864 to 2016 bits per OFDM frame in the $12 \times 12$ situation in \figurename~\ref{fig:speed1_12},
and 1350 to 3150 bits per OFDM frame for the $15 \times 15$ situation
in \figurename~\ref{fig:speed1_15}; where the larger feedback requirement can be ascribed to the presence of more antennas.
It is evident that the lattice based approach is able to track the precoders nearly as well
as the Geodesic and Givens parameter based approaches with a much
smaller feedback requirement (up to 70\% lower than the geodesic and
competitive with the Givens based approaches) with no loss in performance. In fact,
with a similar feedback budget, the proposed lattice filter based channel tracking achieves
higher rates. Although the error, as measured in terms of the Frobenius norm as in \figurename~\ref{fig:ncomp} and
in terms of flag distance as in \figurename~\ref{fig:nflag}, for the proposed approach
seems marginally higher, we note below that this does not affect the
achievable rates significantly.

Due to the gradual variation of the channel, the parameters converge close
to the actual values after approximately 20 adaptation steps. Here, we remark
that the Givens rotation based approach experiences a higher initial error,
since the scalar parameters therein are adapted independently, although it
converges to the minimum error in conjunction with the other approaches. We
remark that, the number of bits fed back to the transmitter is significantly
lower for the lattice structure, compared to Geodesic, Givens, and angle-delay
methods. This can be attributed to the fact that our proposed approach
attempts to capture the precoder characteristics as an all-pass filter
more faithfully, in contrast to the frequency domain methods, which
track precoders at a few frequencies and interpolate the rest, that
could lead to erroneous reconstruction.

Using the adapted lattice parameters, we now compute the achievable
rate as per \eqref{eq3:sysprerate}. The corresponding achievable rates
for different speeds and different MIMO systems are shown in
\figurename~\ref{fig:adapt_rate}. First,
\figurename~\ref{fig:speed1_4} considers a $4\times 4$ system where
with a relative speed of 10~\si{\kilo\meter/\hour}. First, since the
channel variation is low, all methods are able to adapt and track the
channel well. However, we observe a significant gap between in the
achievable rates with the quantised channels and the perfectly
precoded ones. This can be attributed to the fact that, even slight
variations in the precoder could result in inefficient beamforming
that reduces the rate when using linear
receivers~\cite{pitaval2013codebooks}. Nevertheless, we find that that
lattice based approach captures the channel characteristics well, and
is able to nearly match the Givens rotation based channel tracking,
with only 0.3 \si{\deci\bel} separation. In the case of
\figurename~\ref{fig:speed1_8}, we have a situation where the MIMO
system has more antennas ($8 \times 8$), and the speed is also higher
at 50 \si{\kilo\meter/\hour}. Here, we find that the lattice based
approach is able to track the channel much more accurately with fewer
bits than the Geodesic and the Givens methods. This is because the
lattice based parameterisation is able to ``capture'' the 5G channel's
phase (unitary) parameters much better than the other approaches. The
5G channel typically exhibits less of a multipath and more of a strong
line of sight behaviour~\cite{samimi20163} that makes knowledge of
phase more important across frequency, and representation of this
unitary valued phase's variation across frequency is better achieved
by means of a matrix all-pass filter. Further, adaptation using a
smaller number of lattice parameters $\K_k$ (due to the lower order of
the required all-pass filter) is sufficient to achieve rates
comparable to the other methods (12.5\% less feedback than the Givens
approach while outperforming it by about 0.4 \si{\deci\bel}).
\begin{figure}[hbt]
    \centering
    \subfloat[]{\includegraphics[width=\imgwidth]{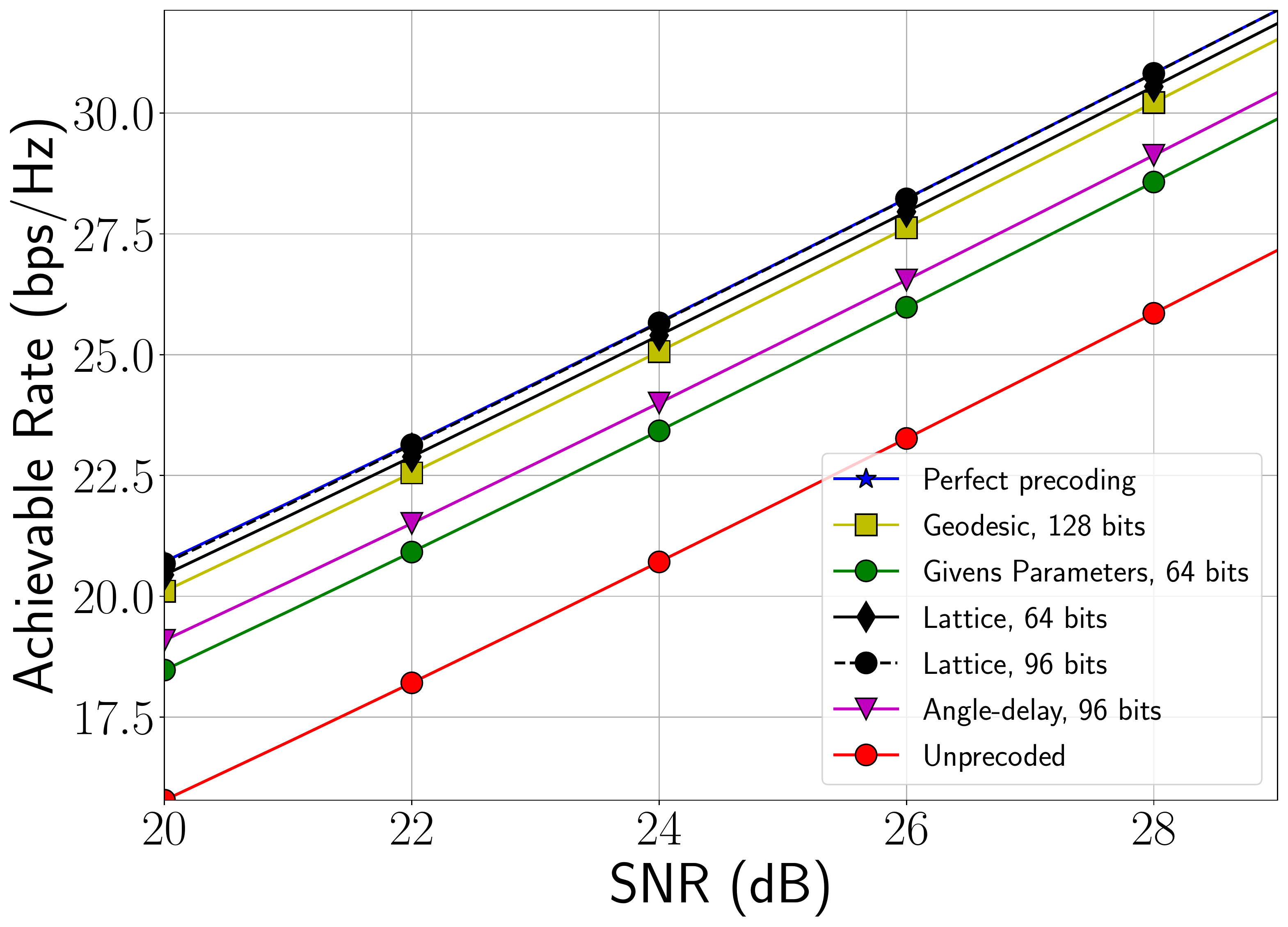}\label{fig:speed1_4}}\\
    \subfloat[]{\includegraphics[width=\imgwidth]{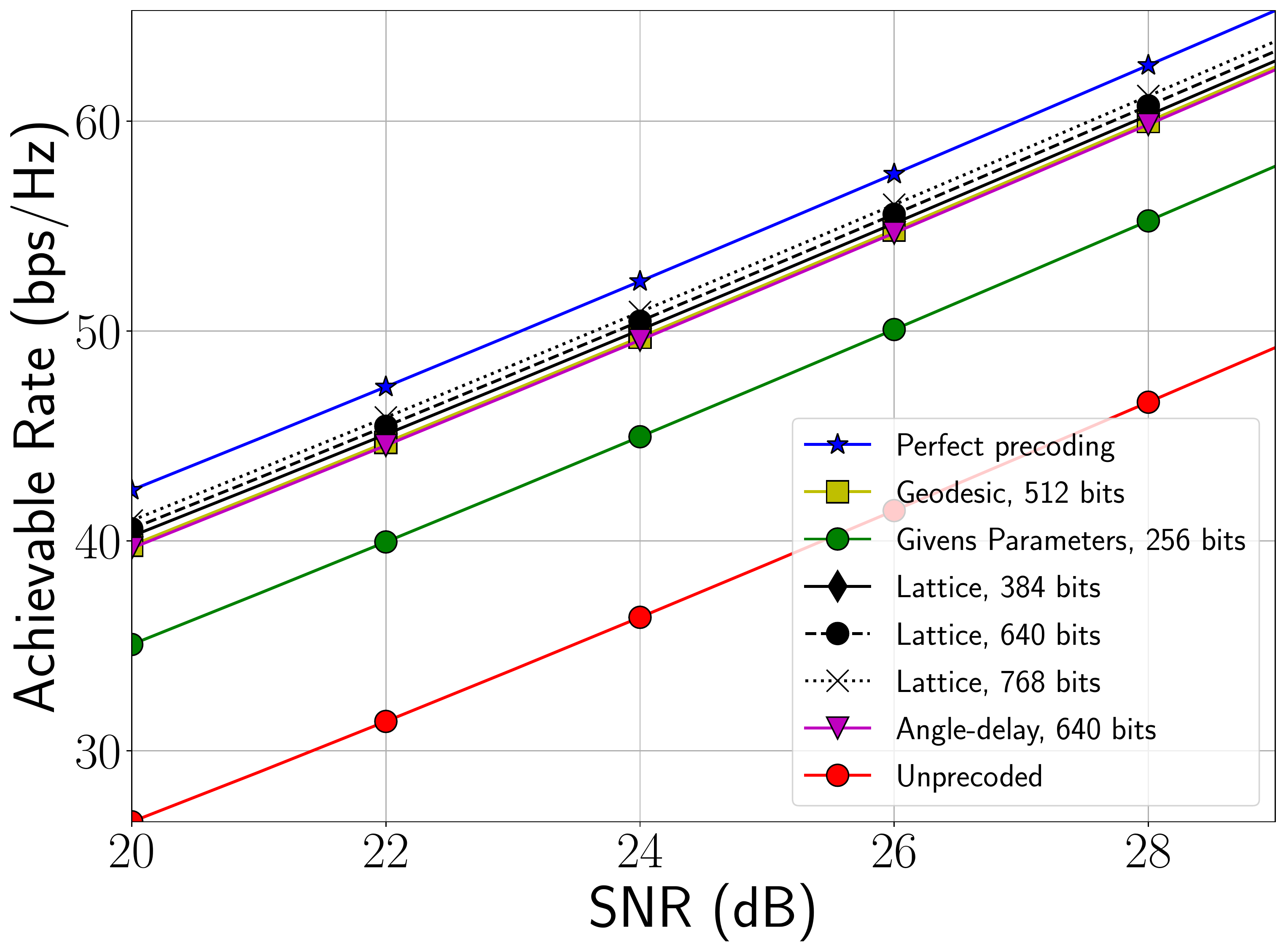}\label{fig:speed1_8}}
    \caption{Comparison of achievable rates between the lattice
      structure and geodesic interpolation approaches for
      \sref{fig:speed1_4} $4 \times 4$ size precoder for relative
      transmitter-receiver speed 10~\si{\kilo\meter/\hour},
      \sref{fig:speed1_8} $8 \times 8$ size precoder for relative
      transmitter-receiver speed 50~\si{\kilo\meter/\hour}.}
\end{figure}
\begin{figure}[hbt]
    \ContinuedFloat
    \centering
    \subfloat[]{\includegraphics[width=\imgwidth]{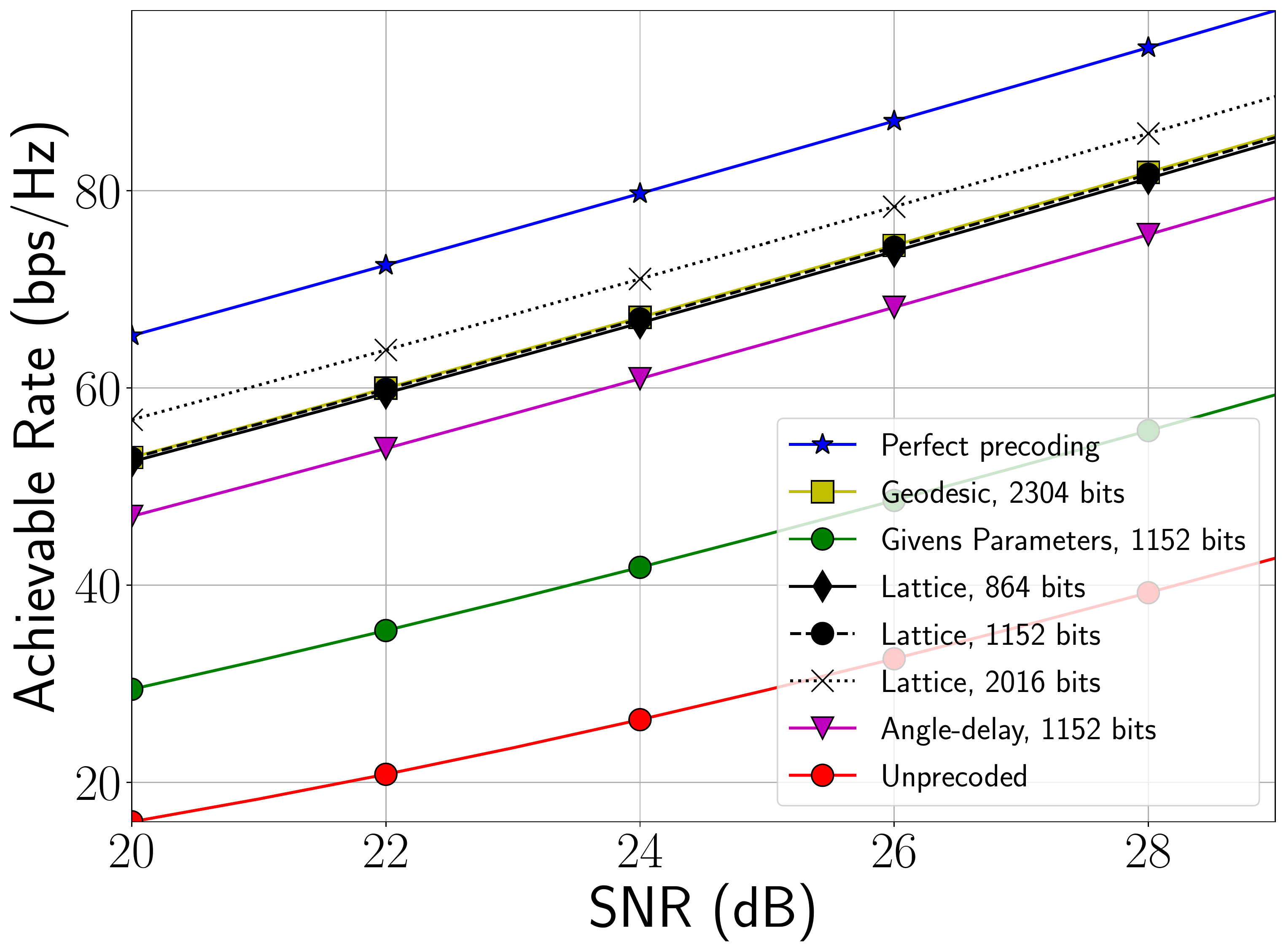}\label{fig:speed1_12}}\\
    \subfloat[]{\includegraphics[width=\imgwidth]{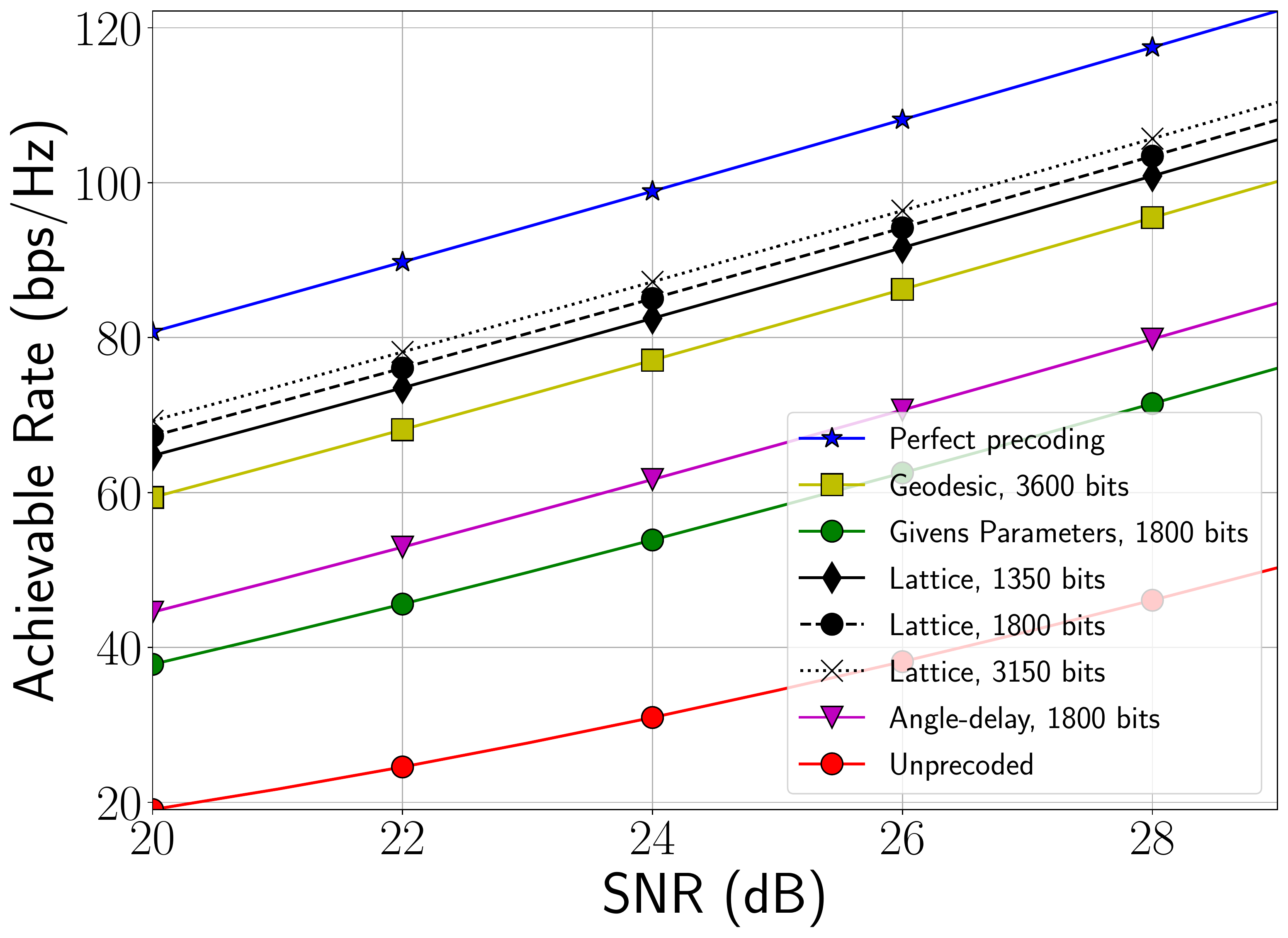}\label{fig:speed1_15}}
    \caption{Comparison of achievable rates between lattice structure and
    geodesic interpolation for \sref{fig:speed1_12} $12 \times 12$ size precoder
    for relative receiver speed 50~\si{\kilo\meter/\hour}, \sref{fig:speed1_15} $15 \times 15$
    size precoder for relative transmitter-receiver speed 100~\si{\kilo\meter/\hour}.}
    \label{fig:adapt_rate}
\end{figure}

For the $12\times 12$ and $15\times 15$ channel sizes, that are more
representatives of the antenna array scenarios in 5G based systems, we
can again confirm the benefits of the proposed approach in
\figurename~\ref{fig:speed1_12} and \figurename~\ref{fig:speed1_15}.
In these situations, since we have a wideband OFDM system, interpolating the
precoders over a wide band of frequencies using both the geodesic as
well as the Givens rotations accrues a significant amount of
error. Thus, at subcarriers that are far away from the pilot locations
where the precoders are estimated and fed back, the error in the
precoder interpolation results in inefficient beamforming, thereby
reducing the rate. However, the reasonably sparse nature of the
channel produces an all-pass filter that can be represented more
accurately using the SNIP approach, even in situations where the
number of antennas is large. Therefore, choosing a much smaller number
of pilot subcarriers, and employing the SNIP is able to capture the
precoder characteristics well even at the points in between the
selected subcarriers. In all of these, tracking $\K_k$ is also
accomplished efficiently, as is confirmed in
\figurename~\ref{fig:speed1_15}.

A similar observation can be made when comparing the performance of
the proposed method with the angle-delay
method~\cite{mo2017channel}.  The angle-delay method uses a truncated
discrete-time response that heavily approximates the overall
precoder. We then quantise and track this approximated response, and
feed it back to the transmitter. In contrast, the proposed method
models the precoder through the coefficients of a matrix IIR all-pass
filter, and further converts them to the lattice structure, that is
suitable for quantisation and tracking while retaining the all-pass
structure. This obviates the need for any time-domain approximations
that are necessary in the case of the angle-delay method.  Therefore,
because of the combined effect of the approximate discrete-time
representation and quantisation, the overall achievable-rate is lower
than that achieved by the proposed method. This is seen
in \figurename{\ref{fig:speed1_4}}, \figurename{\ref{fig:speed1_8}},
\figurename{\ref{fig:speed1_12}}, and \figurename{\ref{fig:speed1_15}}.

To summarise, frequency interpolation
based methods, such as the geodesic and Givens rotation based
approaches produce diminished performance in terms of achievable rate
for larger MIMO systems owing to the larger number of parameters that
need to be interpolated across subcarriers. The SNIP, on the other
hand, tries to match the filter response and fit an appropriate matrix
all-pass filter that yields the desired frequency response with a much
smaller number of interpolating points. Thus, the combined benefit of
efficient representation of the filter and convenient adaptive
tracking makes the SNIP based approach an attractive proposition for
precoder tracking in 5G MIMO communication systems.

\section{Conclusion}
\label{sec:conclude}
In this work we emphasis on lattice based time-domain realisation of a
matrix all-pass filter and present its practicality for designing
precoders for MIMO-OFDM systems. We use the optimised SNIP to obtain
the matrix LCCDE filter coefficients and then convert them to matrix
lattice parameters for convenient representation. By doing so, we
obtain the inherent benefits of the lattice
structure: \emph{stability} and \emph{adaptability}. We show that our
proposed structure is on par with the conventional (frequency-domain)
geodesic and Givens rotation approaches, and angle-delay domain approach
in terms of 1) better
adaptation, 2) comparable achievable data-rates, and 3) smaller number
of feedback bits. Future work would explore more effective techniques
for estimating precoders directly in the lattice form, and the
applicability of the proposed techniques in MIMO joint radar and communication systems.

\bibliographystyle{IEEEtran}
\bibliography{IEEEfull,bibly}

\end{document}